\documentclass[twocolumn, secnumarabic, amssymb, nobibnotes, aps, prd, amsmath, nofootinbib]{revtex4-2}

\newcommand{\bubble}{{\mathrm{b}}}
\newcommand{\metastable}{{\mathrm{m}}}

\newcommand{\muca}{{\mathrm{muca}}}
\newcommand{\canonical}{{\mathrm{ca}}}
\newcommand{\weighted}{w}

\usepackage{graphicx}
\usepackage{hyperref}
\usepackage{xprintlen}
\usepackage{xcolor, algorithm2e}
\usepackage{soul}
\usepackage{mathtools}
\usepackage[mode=buildnew]{standalone}
\begin{document}

\title{Expanded ensemble method for bubble nucleation}
\author{Jaakko Hällfors}
\email[]{jaakko.hallfors@helsinki.fi}
\author{Kari Rummukainen}
\email[]{kari.rummukainen@helsinki.fi}
\affiliation{Department of Physics and Helsinki Institute of Physics,\\ PL64, FI-00014 University of Helsinki, Finland}

\date{Feb 2025}

\begin{abstract}
In the absence of impurities and boundary effects, first order phase transitions are initiated by the nucleation of critical bubbles. In thermally driven transitions many systems can remain metastable for an extended time, possibly tens of orders of magnitude longer than typical microscopic timescales. In standard Markov chain Monte Carlo simulations of these systems the probability of critical bubbles can be too suppressed for the transition to happen in any practical simulation time. The computation can be accelerated by using modified sampling methods, for example multicanonical or Wang-Landau sampling. 
However, even using these methods, there remains a condensation barrier which dramatically reduces the efficiency at large volumes.  We present a novel sampling method, the method of expanded ensembles, which very effectively circumvents the condensation barrier and enables efficient simulations at large volumes.
\end{abstract}

\maketitle

\section{Introduction}\label{sect:introduction}

In the Standard Model of particle physics the high-temperature electroweak phase transition is a smooth crossover \cite{kajantieThereHotElectroweak1996,donofrioStandardModelCrossover2016}. However, a first order electroweak phase transition is possible in many extensions of the Standard Model. In this case the phase transition does not happen immediately as the universe cools down to the critical temperature; the system first supercools and the potential develops a new minimum. Initially the fields are trapped in a metastable phase, separated from the stable minimum by an energy barrier. The phase transition transpires through the nucleation of bubbles of the stable phase. These bubbles then expand and coalesce to complete the transition into the stable phase. Since bubble nucleation is a violent non-equilibrium event, it could provide conditions for a successful baryogenesis \cite{rubakovElectroweakBaryonNumber1996} and generate primordial gravitational waves \cite{capriniDetectingGravitationalWaves2020a, hindmarshPhaseTransitionsEarly2021}.

The planned Laser Interferometer Space Antenna (LISA) should be particularly suitable for detecting gravitational waves originating from first order phase transitions \cite{amaro-seoaneLaserInterferometerSpace2017, capriniScienceSpacebasedInterferometer2016}. Since the goal is to use GWBs for differentiating between the various models, we must be able to predict the spectra accurately. In this work we focus on one of the key parameters, the nucleation rate, which essentially describes how frequently bubbles of the stable phase appear \cite{hindmarshPhaseTransitionsEarly2021}.

Given a prospective model, the nucleation rate can be computed using either perturbative or nonperturbative methods. A classical prescription for the nucleation rate computation was first presented by Langer in \cite{langerStatisticalTheoryDecay1969}. A similar approach for quantum fields at zero temperature was derived by Callan and Coleman in \cite{colemanFateFalseVacuum1977} and \cite{callanFateFalseVacuum1977} and extended to finite temperature by Linde in \cite{lindeDecayFalseVacuum1983}.

For a phase transition at the electroweak scale, the computational problem can be simplified through dimensional reduction; when the temperature is high, heavy scales can be integrated out, leaving us with a three-dimensional effective field theory (EFT). This EFT only contains the bosonic infrared modes, and its thermodynamics are much more amicable to analysis on the lattice \cite{farakos3DPhysicsElectroweak1995b}. Dimensional reduction also transforms the nucleation rate computation into a problem in (almost) classical statistical field theory. In fact, the standard nonperturbative approach is conceptually most similar to Langer's formulation.

Simply put, the lattice computation of the nucleation rate can be factored into a statistical and a dynamical part \cite{mooreElectroweakBubbleNucleation2001}. The statistical factor gives the probability of critical bubbles relative to the metastable phase, while the dynamical factor describes how often the critical bubbles actually evolve into the stable phase. The dynamical factor is computed using Langevin dynamics for the critical bubbles and is in principle easy to perform. The statistical part involves simple Markov chain Monte Carlo (MCMC) sampling of the probability distribution. Due to the extreme probability suppression of the critical configurations interpolating the two phases, multicanonical methods are required.

While the multicanonical method solves the most glaring part of the sampling problem it does not remove all obstacles. Firstly, the periodicity of the lattice results in so-called shape transitions, where the phase interface must morph between different geometric shapes in order to pass from one phase to the other \cite{mooreElectroweakBubbleNucleation2001,neuhaus2DCrystalShapes2003a}. A less obvious and more computationally relevant barrier is given by the condensation transition, where the preferred configuration changes from delocalized fluctuations of the metastable phase into a single localized bubble of the stable phase. In the MCMC computation these transitions appear as barriers to the Markov chain that slow down its exploration of the phase space. Since we need to obtain the probability difference between the regions separated by the barrier, simulations on larger lattices become increasingly arduous. The desire for improved tunneling has been expressed in \cite{gouldFirstorderElectroweakPhase2022}, with improvements presented in \cite{gouldNonperturbativeTestNucleation2024,rummukainenResolvingCriticalBubble2025}. In this work we develop a new method, one that eliminates the condensation barrier and is easily applicable to current systems of interest.

To understand the shortcomings of the standard approach, we begin by describing the sampling problem in Section \ref{sect:multicanonical-method}. We pay particular attention to the effect of the multicanonical weighting to clarify the type of problem it \textit{does not solve}. With these ideas in mind, in Section \ref{sect:tunneling-problem} we investigate the tunneling problem in more detail, briefly with respect to shape transitions, but mostly focusing on the computationally relevant condensation transition and the associated transition barrier. We also consider the effect of using so--called weighted order parameters to manipulate the location of the condensation transition. In Section \ref{sect:expanded-ensemble} we describe the method of expanded ensembles and use weighted order parameters to construct a novel set of subensembles to bypass the condensation barrier. Finally, in Section \ref{sect:toy-model} the effectiveness of the new method is verified for the case of a simple two-dimensional scalar field. We show that our new method eliminates the characteristic exponential slowing down near the condensation transition. Future applications and possible variants of the method are discussed in Section \ref{sect:discussion}.

\section{The nucleation rate and the multicanonical method} \label{sect:multicanonical-method}

The nonperturbative computation of the bubble nucleation rate presents a special type of the sampling problem. Let us consider a system supercooled below the critical temperature of a first order phase transition: the system is initially in the metastable phase and the tunneling to the stable phase is suppressed by the free energy barrier between the phases. The transition to the stable phase happens along the trajectory with the smallest increase in free energy, and the critical bubble is the configuration with the maximum of the free energy on this trajectory, i.e. a saddle point of the free energy landscape.

The energy budget of the bubble consists of the favorable formation of the stable phase and the expense of forming a phase interface. 
Since the former is proportional to volume and the latter to surface area, there exists a critical size (or radius) where bubble growth eventually becomes energetically favorable. This defines the size of the critical bubble. 

Langer's formalism consists of investigating the probability flux through this saddle-point \cite{langerStatisticalTheoryDecay1969}. In particular, the nucleation rate factorizes into a statistical part (statistical probability of the saddle-point configurations) and a dynamical part (the probability flux through the saddle point). Measuring the dynamical part requires a real-time approach, whereas the statistical part can be measured using equilibrium thermodynamics.

For weakly coupled gauge field theories (Standard Model -like) both the statistical and the dynamical part can be measured in lattice simulations \cite{mooreElectroweakBubbleNucleation2001,gouldFirstorderElectroweakPhase2022}. The procedure is essentially equivalent to the earlier nonperturbative sphaleron rate computation of \cite{mooreMeasuringBrokenPhase1998}.

Directing our focus to the statistical factor, let us suppose a system described by a state $s$, and the canonical probability distribution $P_{\canonical}(s) \propto \exp[- H(s)] $. We assume that the system has two phases $A$ (metastable) and $B$ (stable), separated by a region of suppressed phase space. Here we take that an order parameter is a real scalar function $\mathcal O(s)$ such that it obtains distinct values, $\mathcal O_\mathrm{A}$ and $\mathcal O_\mathrm{B}$, in the two phases. The choice of the order parameter $\mathcal O$ is not unique, but a typical candidate usually exists for a given system (e.g. magnetization for the Ising model).
\begin{figure}[htbp]
  \centerline{\includegraphics[]{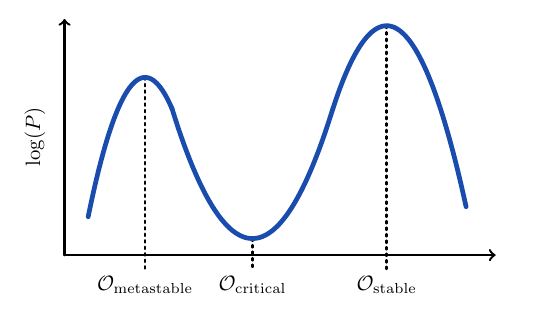}}
  \caption[]{Sketch of $P(\mathcal O)$ below the critical temperature. \label{fig:histogram-structure}}
\end{figure}

As the system traverses from one phase to the other, the order parameter obtains values $O_\mathrm{A} < \mathcal O(s) < \mathcal O_\mathrm{B}$ in the intermediate configurations. This gives $P_{\canonical}(\mathcal O)$ the characteristic two-peaked structure sketched in Fig. \ref{fig:histogram-structure}. The minimum between the peaks at $\mathcal O_\mathrm{critical}$ contains configurations at and near the saddle point. It is precisely the value of $P_{\canonical}(\mathcal O_\mathrm{critical})$ relative to the metastable peak that we seek.

The obvious hindrance to the MCMC sampling of $P_{\canonical}$ results from the fact that the suppression of the bubble configurations can be extreme, even of the order $\sim \mathrm e^{-100}$. Typical MCMC methods will never visit these configurations, since they are not designed to sample suppressed parts of the phase space. The \emph{multicanonical method}, presented first in \cite{bergMulticanonicalAlgorithmsFirst1991}, is however particularly suitable for first order phase transitions. Other modified sampling methods include Wang-Landau \cite{wangEfficientMultipleRangeRandom2001a}, density of states \cite{bennettDensityStatesMethod2024} and the application of Jarzynski's theorem \cite{caselleJarzynskisTheoremLattice2016}. The multicanonical method has been successfully applied to the critical bubble calculation in models related to the Standard Model \cite{mooreElectroweakBubbleNucleation2001,gouldFirstorderElectroweakPhase2022,gouldNonperturbativeTestNucleation2024}, and to SU($N$) gauge theory \cite{rummukainenResolvingCriticalBubble2025}. The method is based on sampling an alternative probability distribution where (optimally) the suppression does not exist.

To elaborate, let us define the \textit{multicanonical probability distribution} for the state $s$ by
\begin{align}
\label{eq:muca-probability-distribution}
  P_{\muca}(s) &\coloneq P_{\canonical}(s) P_{w}(\mathcal O(s)) \\[1ex]
  &\propto \exp[-(H(s) + W(\mathcal O(s)))],
\end{align}
where $W$ is called the \textit{weight function} and depends on $s$ only through $\mathcal O$. $W$ is to be chosen such that the probability distribution of the order parameter in the multicanonical ensemble is uniform. From the form above it is easy to see that the optimal choice is simply to take $P_{w}(\mathcal O) = 1 / P_{\canonical}(\mathcal O)$. The problem, of course, is that $P_{\canonical}(\mathcal O)$ is the unknown function we want to sample. Luckily, a simple iterative process can yield a sufficiently good approximation for $W$.  We detail our method in Appendix \ref{appendix:weight-function-iteration}.

Given a good guess for $W$, our multicanonical probability distribution is now almost flat with respect to $\mathcal O$. The weighting has increased the probability of the intermediate states (relative to the peaks) as to eliminate the canonical probability barrier. To recover expectation values in the canonical ensemble one notes that for any measurable $A$ the canonical expectation value $\langle A \rangle$ can be rewritten as
\begin{align}
  \langle A \rangle  &= \frac{1}{Z} \int \mathcal D s A(s)\exp[-H(s)] \\[1.5ex]
                                 &= \frac{1}{Z} \int \mathcal D s A(s) \exp[W(s)] \exp[- H(s) - W(s)] \\[1.5ex]
                                 &= \frac{\langle A(s) \exp[W(s)] \rangle_{\muca}}{\langle\exp[W(s)]\rangle_{\muca}},
\end{align}
where the multicanonical ensemble should be by construction easy to sample. That is, the canonical expectation value is recovered by weighting each measurement of the multicanonical ensemble with its respective weight. Given a well-behaved $W$ the canonical expectation values obtained through the reweighting are always formally correct. If the choice of $W$ is poor however, obtaining accurate results may still require arbitrarily long runs. A canonical MCMC-computation is easily transformed into a multicanonical one by appending to each regular update the multicanonical accept-reject step.

It is important to think about the choice of $\mathcal O$ carefully, since the effect of the multicanonical method boils down the following statement:
\emph{Using $W(\mathcal O)$ to flatten $P_{\canonical}(\mathcal O)$ corresponds to equalizing the probability mass between canonical ensembles at each fixed value of the order parameter}.

If there exist probability barriers ``orthogonal'' to change in the order parameter  (e.g. multiple metastable states for a fixed value of $\mathcal O$) MCMC sampling of the multicanonical distribution can still be very difficult. In the next section we will see where such barriers can render the standard application of the multicanonical method ineffective.

\section{The tunneling problem in first order phase transitions} \label{sect:tunneling-problem}

\subsection{Shape transitions}

\begin{figure}[htbp]
\centerline{\includegraphics[]{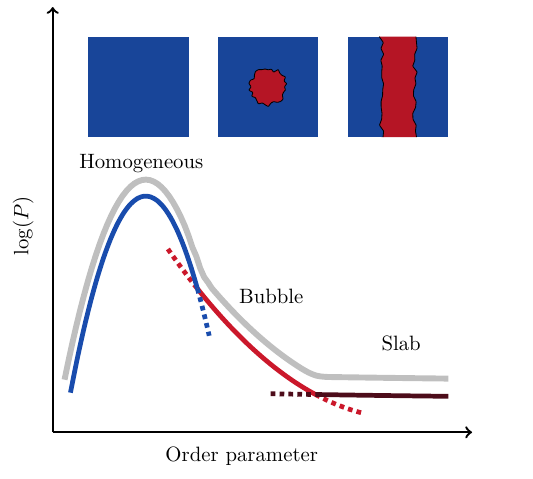}}
\caption{Sketch of the different phase interface shapes for a two-dimensional system. 
In the three-dimensional case there would in addition exist a cylindrical configuration \cite{mooreNonperturbativeComputationBubble2001}. If we were not at the critical value, one of the two peaks would be larger than the other such that the two-peaked distribution would be skewed towards the stable peak. \label{fig:shape-regions}}
\end{figure}

\textit{Shape transitions} give an intuitive example of such orthogonal transitions. In the typical case the simulation volume is taken to be periodic, which leads to distinct minimal energy configurations for the phase interface. For values near the metastable peak the minimal energy configuration is given by a bubble, where a spherical clump of the stable phase is surrounded by the metastable phase. As the value of the order parameter increases, at some point it becomes preferable (for the fixed value of the order parameter) to rather form a cylindrical configuration. We have sketched these regions in Fig. \ref{fig:shape-regions} for a two-dimensional system.

The transition barrier arises from the fact that for the bubble to morph into the cylinder (and vice versa) it must pass through intermediate deformed configurations (here elongated bubbles) of greater free energy cost. The probability of deformed configurations is suppressed for any value of the order parameter and therefore the multicanonical weighting does not remove the transition barrier. These shape transitions happen (in this sense) in an orthogonal direction to change in $\mathcal O$. For our use case they are not relevant, since they are merely undesirable artifacts of the periodic boundary conditions.

\subsection{The condensation barrier}
\begin{figure}[htbp]
  \centerline{\includegraphics[]{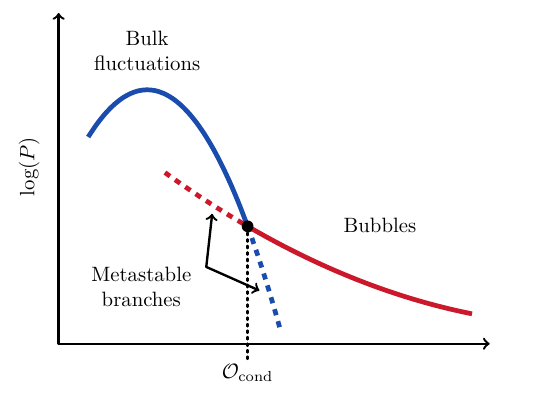}}
  \caption[]{Sketch of the structure of $P(\mathcal O)$ near the condensation transition. The dotted lines describe metastable branches of the bubble and fluctuation configurations. \label{fig:condensation-transition-structure}}
\end{figure}

For the nucleation problem the relevant barrier is given by the condensation transition (see \cite{binderTheoryEvaporationCondensation2003a,binderCriticalClustersSupersaturated1980} for treatment in the Ising model). The condensation transition is located near the value of the order parameter where the preferred configuration changes from delocalized fluctuations of the metastable phase into a single localized bubble of the stable phase. For sufficiently large volumes this shows as a kink in $P(\mathcal O)$, as sketched in Fig. \ref{fig:condensation-transition-structure}. On both sides of the transition point there exist metastable branches for both small bubbles and large bulk fluctuations \cite{binderTheoryEvaporationCondensation2003a}. In Fig. \ref{fig:condensation-transition-structure} the critical bubble would be at the minimum of the bubble branch off to the right. For a poor choice of order parameter this is not necessarily the case; the critical bubble can be smaller than the bubble at the condensation transition, in which case the critical bubbles can not be identified. In this work we presume that this is not a problem.

It is now convenient to imagine a measurable $\lambda$ that could distinguish whether a configuration contains a bubble or not. This has been constructed for the two-dimensional Ising model \cite{nussbaumerFreeEnergyBarrierDroplet2010}, but we will keep the discussion on a heuristic level. The probability distribution of $\lambda$ in the transition region would be bimodal, with peaks corresponding to the bubble and fluctuation branches. At the condensation transition these peaks are of equal volume, while near it one of the peaks, that corresponding to the stable branch, dominates.

\begin{figure}[htbp]
  \centerline{\includegraphics[]{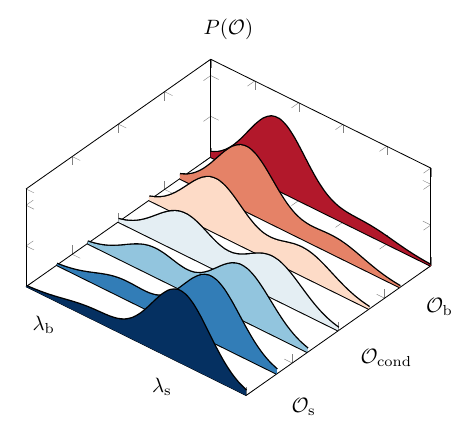}}
  \caption[]{Sketch of $P(\mathcal O)$ near the condensation transition with the hidden metastability in the direction of $\lambda$. Subscripts b and s denote the bubble and metastable branches respectively. \label{fig:muca-hidden-barrier}}
\end{figure}

\begin{figure}[htbp]
  \centerline{\includegraphics[]{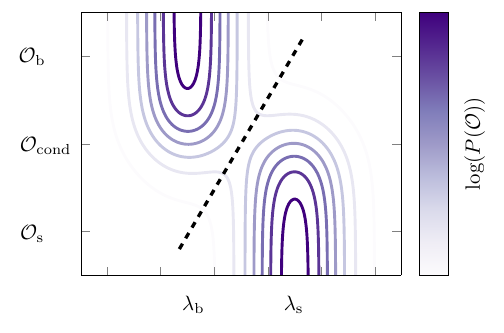}}
  \caption[]{The probability distribution of Fig. \ref{fig:muca-hidden-barrier} from above. The dashed black line indicates the location of the condensation barrier in the multicanonical ensemble. If there existed an order parameter that varied perpendicular to the dashed line, the corresponding multicanonical simulation should have no trouble smoothing out the barrier.\label{fig:hidden-barrier-contour}}
\end{figure}

To see why the barrier exists in the multicanonical ensemble, we have sketched the transition region in Fig. \ref{fig:muca-hidden-barrier}. As we recall from earlier, the effect of the multicanonical weighting is to equalize the ``area'' under the probability distribution $P(\lambda)$ for each value of the order parameter. As the condensation transition happens in the direction of changing $\lambda$, this multiplicative weighting does not affect the bimodal nature of $P(\lambda)$ and the barrier between the two branches remains. This is further illustrated in Fig. \ref{fig:hidden-barrier-contour}, where we sketch probability distribution of Fig. \ref{fig:muca-hidden-barrier} from above. We emphasize that in the context of field theories the present discussion on the condensation transition is a heuristic one, based on analogous results in the context of the two-dimensional Ising model \cite{binderTheoryEvaporationCondensation2003a,nussbaumerMonteCarloStudy2008a,nussbaumerFreeEnergyBarrierDroplet2010}.

The height of the barrier is essentially given by the suppression between the peaks of $P(\lambda)$ at $\mathcal O_\mathrm{cond}$. If the physical size of the corresponding bubble at $\mathcal O_\mathrm{cond}$ is large, the intermediate states between the branches become increasingly suppressed. Therefore for the MCMC computation it is desirable to have the condensation transition occur at the smallest possible physical bubble size. To this end, the choice of order parameter is essential. It would be preferable for the order parameter to have a small variance in the metastable phase\footnote{This is to be taken relative to the values the order parameter takes in the two phases. E.g. simple scaling will not have an effect as it does not change the foliation of the phase space in a nontrivial manner.}, since this results in the condensation transition occurring at smaller physical bubble size.

In this sense it is sometimes said that an order parameter with small variance can ``see'' smaller bubbles. Of course, for larger system volumes there exists more bulk phase and therefore more bulk fluctuations, resulting in the condensation transition shifting to ever larger physical bubble size.

One way to alleviate this problem is to find alternative order parameters of low variance with examples constructed in \cite{gouldNonperturbativeTestNucleation2024,rummukainenResolvingCriticalBubble2025}. More complicated measurables may also be considered, but they can become too computationally expensive for use in the multicanonical weighting or otherwise difficult to interpret, possibly foliating the phase space in a manner that inefficiently locates the critical bubbles. In addition, it would be a tedious affair to invent such order parameters for every system of interest. We will now show how the typical order parameter can be used to construct a set of related low-variance order parameters.

\subsection{Measuring in subvolumes} \label{subsect:measuring-subvolumes}
Consider what happens when we only measure the order parameter in some subvolume of the lattice. We assume the following usually attainable properties of the order parameter: it is able to resolve a large bubble on the lattice and is a volume average of some local measurable of the field values $s_{i}$. That is, it can be written as
\begin{equation}
	\mathcal O(s) = \sum_{i}f(s_{i}),
	\label{eq:mean-OP-type}
\end{equation}
where $f$ is a function of the value $s_{i}$ at site $i$, and possibly some of its nearby neighbors. When the given by this form, it usually makes sense to measure the order parameter in some well-behaved (e.g. cubical or spherical) subvolume of the lattice. These modified measurables will be known as \emph{weighted order parameters}.

The values of a weighted order parameter only depend on the state of the system in the subvolume considered. This results in some useful features. Heuristically, the probability distribution $P(\mathcal O^{\weighted})$ of a weighted order parameter $\mathcal O^{\weighted}$ (in a subvolume $V^{\weighted} \subset V$) should quite closely match that of $\mathcal O$, provided that $\mathcal O$ is considered in a system of volume $V^\prime \approx V^{\weighted}$ (boundary effects aside). Most importantly, the condensation transition occurs at a similar physical bubble size as in a system of volume $V^\prime$. Through a choice of small $V^\weighted$ it is then possible to ``see'' smaller bubbles (within the subvolume) than when using the regular order parameter.

When a weighted order parameter is used for the multicanonical weighting, deviations from the metastable minimum are only encouraged within $V^{\weighted}$. Furthermore, for sufficiently large values of $\mathcal O^{\weighted}$ the system is almost certainly in the bubble branch of $P(\mathcal O^{\weighted})$. If $\mathcal O^{\weighted}$ is restricted to a suitable interval, the bubbles on the bubble branch should fit inside the subvolume and be similar to those found in the regular multicanonical ensemble. As the condensation transition of the weighted ensemble occurs at small physical bubble size, the corresponding condensation barrier is also small and the transition between the two branches of $P(\mathcal O^{\weighted})$ should be essentially unhindered.

Of course, this alone does not solve the initial problem of obtaining the probability of large bubbles. In the typical problematic scenario, no particular choice of subvolume $V^{\weighted}$ will be small enough to reduce the condensation barrier for the corresponding $\mathcal O^{\weighted}$ while also being large enough to fit critical bubbles of desired size. This is where the method of expanded ensembles comes in: we can use different subvolumes in different parts of $P( \mathcal O)$ in a manner that results in the condensation barrier being restricted only to the small and unproblematic subvolumes.

\section{Method of expanded ensembles} \label{sect:expanded-ensemble}
Introduced in \cite{lyubartsevNewApproachMonte1992}, the method of expanded ensembles is most commonly found in the form of the familiar simulated tempering, where one considers additional subensembles that vary in their temperature. In general, however, the subensembles can use near arbitrary probability densities. To elucidate the usefulness of this property in the context of our multicanonical simulations, consider as a starting point the now familiar system from Section \ref{sect:multicanonical-method} described by the multicanonical probability density
\begin{equation}
	P(s) \propto \exp [- \{H(s) + W(\mathcal O(s))\}].
\end{equation}
The problem of $P(s)$ was the fact that the multicanonical weighting, based on the full-volume order parameter, is unable to efficiently encourage the formation of bubbles.

Using the expanded ensemble method, we can instead consider $N$ multicanonical probability distributions $\{P(s, i)\}$ with
\begin{equation}
  P(s, i) \propto \exp\left[-\{H(s) + W^{\vphantom w}_i(\mathcal O^w_i(s)) + \eta_i\}\right],
\end{equation}
where $\mathcal O^w_i$ are different order parameters, $W^{\vphantom w}_i$ their respective weight functions, and $\eta_i$ are constants. Now, in addition to the regular updates modifying the configuration $s$, we can also propose to change the subensemble from $i$ to $j$. A symmetric proposal is easy to fashion with, say,
\begin{equation}
    q(i,j) =
  \begin{cases}
    1/2 & \mathrm{for\ } j = i \pm 1 \\
    0   & \textrm{otherwise}
  \end{cases}.
\label{eq:proposal}
\end{equation}
The resulting acceptance probability for the ensemble update is then simply
\begin{align}
  \begin{split}
  &\hspace{1mm}\mathcal P(i \rightarrow j) \\[1ex]
  & = \min \left(1, P(s, j) / P(s, i)\right)
  \end{split}\\[1ex]
  \begin{split}
                              & = \min (1, \exp [- \left(W^{\vphantom{w}}_j (\mathcal O^{\weighted}_j(s)) - W^{\vphantom{w}}_i (\mathcal O^{\weighted}_i(s))\right)  \\[0.5ex]
                              & \hspace{2.1cm} - (\eta_j - \eta_i)]).
  \end{split}
\end{align}
The constants $\eta_{i}$ should be chosen such that this acceptance probability is on average symmetric.

To see the merit in this ordeal, recall the weighted order parameters from earlier. We reasoned that these are much more effective at creating small bubbles than the full-volume order parameter. The idea is then to use the weighted order parameters to construct additional multicanonical ensembles where bubbles are easily created and destroyed. The ensemble update can then feed these configurations into the original multicanonical ensemble, essentially removing need to cross the condensation barrier. The actual process of constructing these subensembles is only slightly more involved and will be described next.

\subsection{Choosing subensembles for the condensation barrier}\label{subsect:subensembles-for-condensation-barrier}
\begin{figure*}[htbp]
  \includegraphics[width=\textwidth]{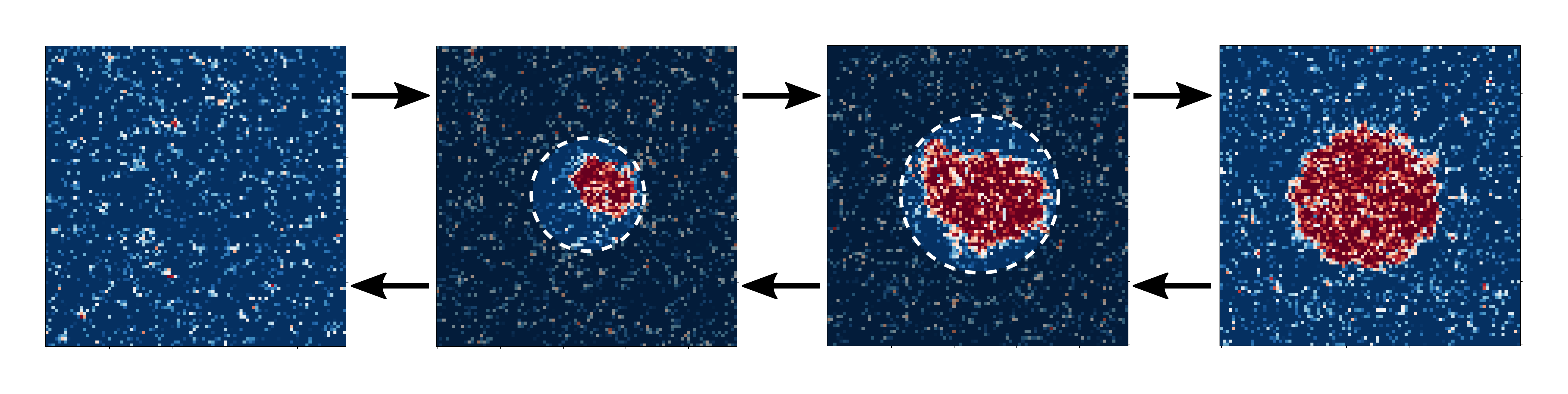}
  \caption[]{Sketch of the subensembles of the expanded ensemble. The light areas mark the sites used in the computation of the weighted order parameter. The first and last subensembles compute the order parameter in the full volume. Values near the metastable peak are colored blue, while the red blob represents a bubble of the stable phase.
    \label{fig:subensembles}}
\end{figure*}

The subensembles $P(s, i)$ will be multicanonical ensembles using different weighted order parameters. We present the subensembles as ordered (in the index $i$), since they are only designed to overlap with their ``neighboring'' subensembles. An illustration of the subensembles is shown in Fig. \ref{fig:subensembles}. The first and last subensemble use the regular full-volume order parameter, and are restricted to the metastable peak and bubble branch respectively. These are the subensembles of interest, since they sample parts of the original probability distribution $P(\mathcal O)$. The goal of the Monte Carlo computation is to recover the relative probability difference between these two subensembles, as it is the quantity that the standard multicanonical method is unable determine efficiently due to the condensation barrier. The intermediate subensembles are too constrained to correspond to physical ensembles (translation invariance is broken) and only serve as a bridge connecting the physically interesting endpoints.

We choose the first weighted order parameter such that the condensation barrier for the given subvolume is small. The values of this weighted order parameter are restricted onto an interval so that the largest value corresponds to a bubble that still fits inside the subvolume. This is to prevent deformed bubbles; To obtain the overlap between the subensembles, we want the largest bubbles of the preceding subensemble to be the typical small (relative to the subvolume) bubbles of the next one. For the first weighted subensemble the lower limit can be taken as the location of the metastable peak, $\mathcal O_{\metastable}$, though this is not particularly important.

For the next subensemble we choose the subvolume to be some constant factor larger than the preceding one. We choose the upper limit again so that the bubbles fit inside subvolume, while the lower limit is chosen such that the small bubbles (relative to the subvolume) of this subensemble are of similar size as the largest bubbles of the preceding one. This is not a very precise statement and it appears sufficient to estimate the lower limit using the value of the order parameter in the metastable phase. That is, we estimate the next interval from the previous one by assuming that outside of the weighted subvolume the system is in the metastable phase. This yields the estimate
\begin{equation}
  \label{eq:full-volume-estimate}
  \mathcal O^{w}_{n + 1} V^{\weighted}_{n + 1} \approx \mathcal O^{w}_{n} V^{\weighted}_{n} + \mathcal O_{\metastable} (V^{\weighted}_{n+1} - V^{\weighted}_{n})
\end{equation}
and therefore to obtain an overlap of $q \in \left( 0, 1 \right)$ the lower limit is
\begin{equation}
	\mathcal O^{w, \mathrm{low}}_{n + 1}  = \left(q \mathcal  O^{w, \mathrm{low}}_{n} + (q - 1) \mathcal O^{w, \mathrm{high}}_{n} \right) \dfrac{V^{\weighted}_{n}}{V^{\weighted}_{n + 1}}.
	\label{eq:subensemble-limits}
\end{equation}
For example, $q = \frac{1}{2} $ should yield a lower limit that (approximately) corresponds to the midpoint of the previous subensemble with respect to the full-volume order parameter. We use $q = \frac{2}{3}$ for our examples, but stress that this choice has no particular significance.

The process is repeated until the subvolumes approach the system volume. The last subensemble, now again using the full-volume order parameter, is finally attached to the end of the chain. Note that the upper limit for the weighted order parameter can usually be set to a constant for all subensembles, since it corresponds to setting the upper limit for the  ratio $V^{\weighted}_\bubble / V^{\weighted}_i$. This is the quantity that determines whether the bubble fits inside a given subvolume and is effectively independent of $V^{\weighted}_{i}$. Furthermore, it is easily estimated from the histogram of a standard multicanonical simulation performed at small, manageable volume. This iterative determination of the subensembles is easily automated and requires relatively little information of the system at hand.

As a result the weighted order parameters following the first one are always restricted to their respective bubble branches. That is, the condensation barrier only exists in the first weighted subensemble where the small subvolume guarantees that the condensation barrier is weak and \textit{independent of the full volume.} Due to the imposed limits, in the other weighted subensembles the bubbles can only shrink and grow, not evaporate.
In practice, depending on e.g. the overlap parameter $q$ and the location of the condensation transition on the order parameter interval of interest, more than one weighted subensemble might contain the transition barrier. Choosing sufficiently small subvolumes for the first few subensembles still results in the barrier being negligible.

Only few quantities need to be known in order to apply the method: the value of $\mathcal O$ in the metastable phase, $\mathcal O_{\metastable}$, and some upper limit for $\mathcal O^{\mathrm{max}}$ which ensures that bubbles fit inside the chosen subvolumes. From these it is easy enough to construct the limits for the unweighted subensembles as well. The initial subvolume, the factor of volume increase between subvolumes, and the overlap parameter appear to require little to no tuning.

This simple construction comes with a small caveat: the subensembles should only contain bubbles well-below the volume of the critical bubble. If this volume is exceeded in the weighted subensembles the expanding bubble can slip outside of the subvolume (since bubble growth is now the energetically preferable direction) and the system can end up entirely into the stable phase, leaving the Markov chain stuck. This could be avoided by adding further boundaries to the weighted ensembles that prevent the full-volume order parameter from going past the minimum.

Fortunately, this is typically not a problem since we only need to find the probability difference between the metastable peak and any part of the well-resolved bubble branch. The rest of the branch (including the critical value) is easily investigated using the standard multicanonical method. This result can then be simply glued onto the previously obtained section of the bubble branch. Since the critical bubble is not hidden by the bulk fluctuations (in a sensible setup) the subcritical part of bubble branch is always available.

\section{A practical example: the two-dimensional scalar field}\label{sect:toy-model}
To demonstrate the method we consider a simple scalar field $\phi$ on a periodic two-dimensional lattice with the discretized Hamiltonian
\begin{equation}
	\label{eq:2D-action}
	H(\phi) = \sum_{x} \sum_{\hat \mu} \left( \phi (x + \hat \mu) - \phi(x)  \right)^{2} + \sum_{x} V(\phi(x)),
\end{equation}
where
\begin{equation}
  \label{eq:2D-potential}
  V(\phi) = \lambda \phi^{4} - m^{2} \phi^{2}
\end{equation}
and the corresponding probability density is given by
\begin{equation}
  \label{eq:2d-prob-density}
  P( \phi) \propto \exp[-\beta H(\phi)].
\end{equation}
The parameters are chosen so that the expectation value of $\phi$ is non-zero:
\begin{equation}
  \frac{1}{V} \hspace{0.5mm}\bigg\langle \sum_x \phi(x)  \bigg\rangle  = \pm v \ne 0.
\end{equation}
Let us consider the first order phase transition from $\langle \phi\rangle = +v $ to $-v$.
As the potential is symmetric with respect to $\phi \rightarrow -\phi$, the system is effectively at the critical point and the critical bubble is infinitely large. This is not essential for the argument for it is sufficient to obtain the probability density on the subcritical part of the bubble branch.  $\langle \phi \rangle$ is a natural choice for the order parameter and when used for multicanonical weighting the condensation barrier becomes apparent even for small volumes. Schematically, the order parameter probability distribution is similar to that we presented in Fig. \ref{fig:shape-regions}.

\begin{figure*}[htbp]
    \centerline{\includegraphics[scale=1]{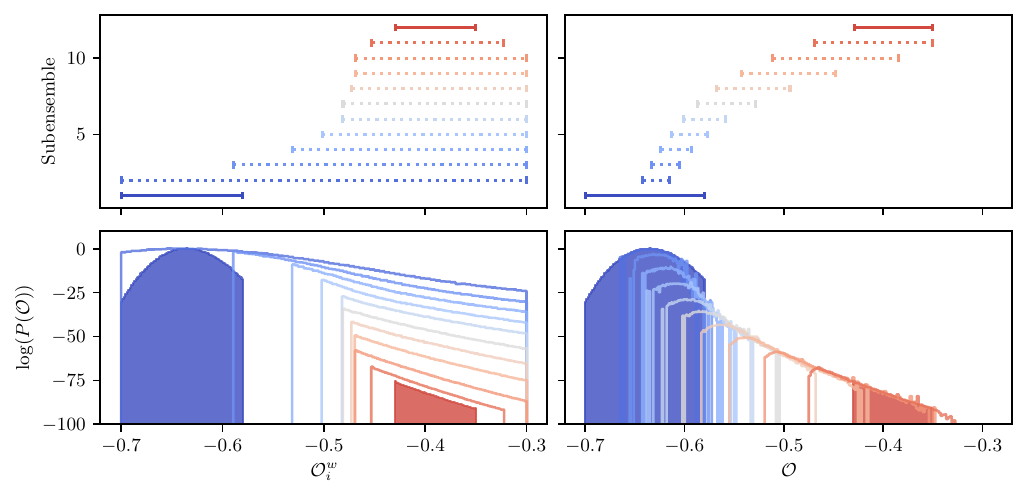}}
    \caption[]{On the left (right) we show the weighted (full-volume) order parameter intervals and the resulting probability distributions from a run with $L = 24$. The solid lines in the upper panels and filled histograms in the lower panels correspond to the full-volume subensembles. Note that the intervals for the full-volume order parameter are estimates based on Eq. \ref{eq:full-volume-estimate}. \label{fig:subensemble-intervals-histograms}}
    \end{figure*}

We take the parameters to be $\beta = 5$, $m=1$ and $\lambda=1$, because this results in a condensation barrier that increases conveniently fast as a function of the system volume. This is a simple balancing act: if the barrier is too strong only a small range of system volumes are feasible using the standard multicanonical method. On the other hand, if it is too weak, very large lattices are needed in order to observe the resulting slowing down. We note that the parameters are not implied to carry any particular physical significance or relation to the continuum limits usually desired in a typical lattice field theory computation. The simulation program was written using the HILA lattice simulation framework \cite{HILALatticeSimulation}.

We perform Monte Carlo simulations of the model using standard multicanonical method and the method of expanded ensembles. For both methods the fields are updated by proposing Gaussian changes $\delta \phi$ on even and odd lattices sites with the standard Metropolis-Hastings algorithm and a multicanonical accept-reject step. The expanded ensemble simulation also performs an ensemble swap update and a shift update. The shift update is used to counter the breaking of translation invariance in the weighted subensembles and consists of a simple spatial shift of the field configurations. All in all, one update sweep then consists of the steps
\begin{itemize}
  \itemsep0.2em
\item[(a)] Multicanonical update of the field configuration
\item[(b)] Ensemble swap update
\item[(c)] Shift update
\end{itemize}
The standard method only uses step (a). 25 such sweeps are performed between each measurement.

The shift proposal is generated from uniform distribution into directions $\{\hat x, - \hat x, \hat y, - \hat y\}$. The magnitude of the shift is obtained as values drawn from $U([0, L / 2])$ rounded to the nearest integer. Again as $H(\phi)$ is unchanged (translation invariance), the update is accepted with the probability
\begin{equation}
  P_{\rm shift} = \exp[-(W^{\vphantom{\weighted}}_i(\mathcal O^\weighted_{i}(\phi')) - W^{\vphantom{\weighted}}_i(\mathcal O^\weighted_{i}(\phi)))],
\end{equation}
where $\phi'$ is the linearly shifted $\phi$. This update is trivial in the first and last subensembles using the translation invariant full-volume order parameter.

For the weighting we use spherical subvolumes with the initial weight radius as $r_{\mathrm{init}} = 5$ and obtain successive weight radii through multiplication by $\sqrt{4/3}$. This corresponds to about $1/3$ increase in $V^{\weighted}$ on the lattice (deviations result from the fact that the lattice is discrete). It must be emphasized that these numbers carry no particular significance to the specific system at hand. Perhaps the only important requirement is that the regular multicanonical simulation with $V \approx V^{\weighted}_{\mathrm{init}}$ should face a negligible transition barrier for the expanded ensemble method to be effective. The subensemble intervals resulting from this procedure for $L = 24$ are drawn in Fig. \ref{fig:subensemble-intervals-histograms}, along with the histograms from the resulting run.

\begin{figure}[htbp]
\includegraphics[scale=1]{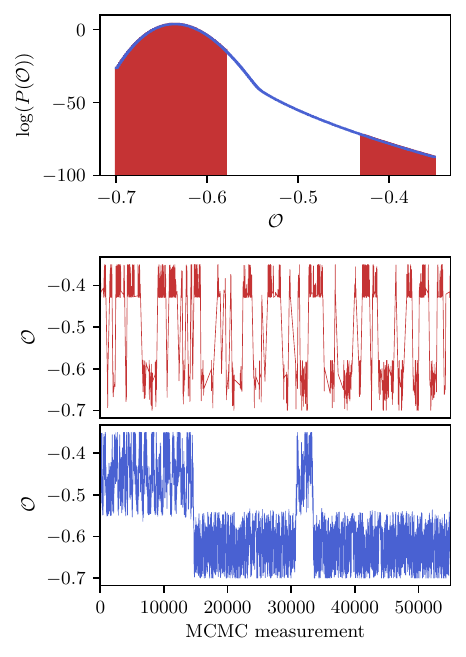}
\caption{In the topmost panel we show how the histograms from the two methods match for $L = 32$ (filled histogram is for the expanded ensemble). The two bottom panels have sections of the corresponding Markov chains. For the expanded ensemble only measurements corresponding to the first and last ensembles have been included. \label{fig:2D-muca-ee-histogram-comparison}}
\end{figure}

In Fig. \ref{fig:2D-muca-ee-histogram-comparison} we show the comparison between the Markov chains and obtained probability distributions for $L = 32$. It is easy to see that the tunneling rate between the two parts of the distribution is much improved using the expanded ensemble method. Additionally, as should be, the resulting probability distributions are in good agreement where they overlap.

\begin{figure}[htbp]
  \includegraphics{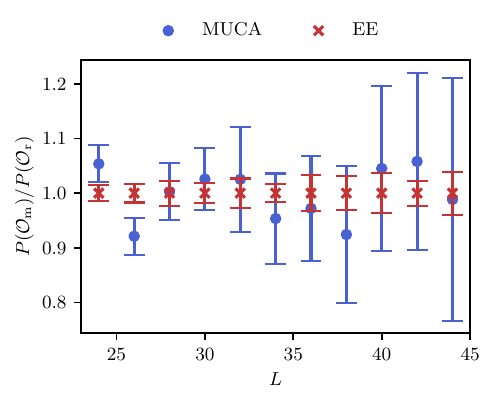}
  \caption{Probability differences and their error estimates normalized with respect to the expanded ensemble method. The results agree, but the statistical uncertainties in the multicanonical simulations grow rapidly as the volume increases.\label{fig:2D-muca-ee-value-comparison}}
\end{figure}

We also compare the probability distributions in more detail in Fig. \ref{fig:2D-muca-ee-value-comparison}, where we have computed the ratio of the probability densities on both sides of the barrier. In particular we compare the values at the metastable peak $\mathcal O_{\metastable}$, and at $O_{\mathrm{r}} = -0.4$ such that the desired ratio is given by $\delta P = P(\mathcal O_\metastable) / P(\mathcal O_\mathrm{r})$. All data has been normalized with respect to the values from the expanded ensemble method. The error bars are obtained from simple jackknife analysis with each chain divided into ten sections. Note that since $\mathcal O$ is a volume average, same numerical value for different lattice sizes corresponds to different sized bubble and therefore different relative probability. We see that both methods agree well with the expanded ensemble method also yielding much reduced error estimates.

To quantify the improvement in the tunneling rate we must first define the tunneling event. As the description of the probability distribution consisting of a metastable peak and a bubble branch is not a strict one for finite volume, we construct a simple heuristic definition. First we arbitrarily choose two regions based on the full-volume order parameter: one for the metastable peak and another for the bubble branch. To the metastable peak belong values with $\mathcal O < \mathcal O_\metastable$ and to the bubble branch $\mathcal O_{\bubble} < \mathcal O$.

\begin{figure}[htbp]
\centerline{\includegraphics[]{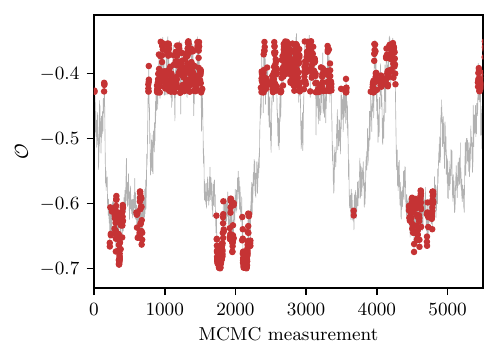}}
\caption[]{\label{fig:2d-ee-chain-visualisation} Section of the Markov chain of $\mathcal O$ for $L=32$ in the expanded ensemble. The gray line includes all subensembles and we have used red dots to highlight those that are in either the first or last subensemble.}
\end{figure}

\begin{figure}[htbp]
  \includegraphics{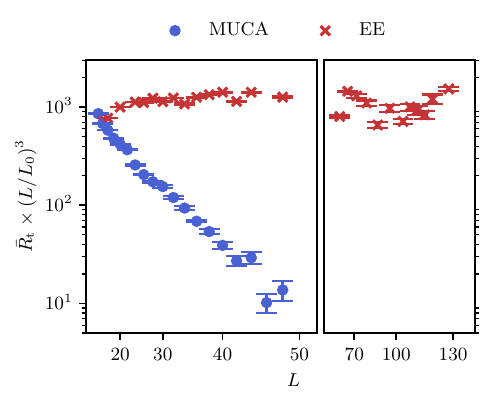}
  \caption{Appropriately scaled normalized tunneling rate for different two-dimensional lattice sizes. The $L$-axis is scaled as $L^{2}$, the system volume. In addition we have split the horizontal axis in order to fit the largest volumes. This results in the two panels having different horizontal scaling. We have picked the threshold values $\mathcal O_{\metastable} = -0.63$ and $\mathcal O_{\bubble} = -0.41$ to define a crossing event. The error estimates are based on ten jackknife samples. \label{fig:2D-muca-ee-tr-comparison}}
\end{figure}

The Markov chain is to hold a label that indicates the region it has visited last, with the change in label counted as a tunneling event. This definition works for both methods when we additionally require that the Markov chain of the expanded ensemble is in the first or last subensemble (those that sample the same probability distribution as the standard method). We have illustrated this in Fig. \ref{fig:2d-ee-chain-visualisation}, where we display the full Markov chain highlighting the points that are in the two physically interesting subensembles.
We define a normalized tunneling rate $\bar R_{\mathrm{t}}$ as the number of tunneling events ($N_{\mathrm{t}}$) per $10^{8}$ MCMC update sweeps.
In Fig. \ref{fig:2D-muca-ee-tr-comparison} we show the normalized tunneling rate $\bar R_{\mathrm{t}}$ for the expanded ensemble method and the regular multicanonical method. We have multiplied both $\bar R_{\mathrm{t}}$:s by $\propto V^{3/2}$ to factor out some of the volume scaling. We expect that the in the absence of the condensation barrier the rate should be $\propto V^\alpha$. With the chosen scaling the normalized tunneling rate of the expanded ensemble method is consistent with constant while the standard multicanonical method suffers from severe exponential slowing down.

We note that $\bar R_{\mathrm{t}}$ does not vary very smoothly as a function of L. Aside from statistical errors, there are always some variations in the iterated weight function. We have checked that the multicanonical distributions for all our runs are approximately flat (with the probability density between the two branches differing by at most a factor of $2$). For the expanded ensemble these variations are expected since the number and limits of the intermediate ensembles depend on the full system volume. Additionally, the number of tunneling events is relatively small for the larger systems ($N_{\mathrm{t}} \approx 150$ for $L = 128$).

Na\"{i}ve extrapolation of the tunneling times of the standard multicanonical method suggests that our improved procedure is able to investigate system volumes that would otherwise be completely out of reach.

\section{Discussion}\label{sect:discussion}

We have presented a novel combination of the multicanonical and expanded ensemble methods for the nonperturbative determination of the bubble nucleation rate. Based on the results obtained for the toy model, the condensation barrier faced by the standard multicanonical method is eliminated. That is, with the new method the tunneling times do not increase exponentially with the system volume, but are reduced to a relatively mild scaling, consistent with $\propto V^{3/2}$. The new method is trivially generalized for the three-dimensional case and the typical order parameters used in many systems of physical interest. Preliminary testing suggests that sampling of the critical bubbles is also sped up in the cubic anisotropy model (see \cite{mooreNonperturbativeComputationBubble2001} for the standard method), although there the exponential slowing down is less apparent for the range of volumes investigated so far. 

For future applications, provided that the condensation barrier can now be efficiently circumvented, the problem of obtaining the statistical probability of very large bubbles becomes much more approachable. Such nucleation rate computations at low supercooling would allow for an interesting comparison to the typical thin-wall approximation. Unfortunately the two-dimensional toy model of this work does not yield useful results, since the lattice spacing is most likely too large for modeling the finite thickness of the phase interface.

Although of limited use in the nucleation rate computation, it would be interesting to investigate whether weighted ensembles could also be used for bypassing shape barriers discussed earlier. Through the use of appropriately deformed weighted subvolumes it may be possible to create subensembles where the otherwise unlikely deformed phase interfaces are preferred. These could then efficiently interpolate between the different geometric configurations.

\subsection*{Acknowledgments}

We acknowledge support from the ERC grant CoCoS 101142449, and the Research Council of Finland grant
354572.

\appendix

\section{Implementation of the algorithm} \label{sect:computational-implementation}
The probability density of expanded ensemble takes the form
\begin{equation}
	P \left( s, i \right) = \exp\left[ - H(s, i) \right],
	\label{eq:weighted_OP_EE_definition}
\end{equation}
where
\begin{equation}
	H(s, i) = H(s) + W^{\vphantom{\weighted}}_{i}\left( \mathcal O_{i}^{\weighted} \left( s \right) \right) + \eta_{i},
	\label{eq:hamiltonian_parts}
\end{equation}
$H(s)$ being the original Hamiltonian of interest. First the weight functions $W_{i}$ are iterated separately for each subensemble. After this the relative weights $\eta_{i}$ are iterated with the subensemble weight functions fixed. Separating the weight function from the relative weights is inconsequential but convenient as it is then easier to estimate the time taken by the weight iteration process. The first round of weight iteration tunes the subensemble histograms flat and the relative weights aim to do the same for the distribution of the index $i$. The iteration method is detailed in Appendix \ref{appendix:weight-function-iteration}.

\section{Iteration of the multicanonical weight function} \label{appendix:weight-function-iteration}
As the probability distribution $P(\mathcal O )$ is a priori unknown, the multicanonical weight function $W(\mathcal O)$ must be obtained through an iterative process. Plenty of methods exist, but we have resorted to a very simple heuristic approach. Note that during the actual MCMC computation the weight function is fixed and the simulation is formally correct for any finite $W$. That is, $W$ only needs to be a good enough approximation. This can be verified afterwards by the observation that the resulting multicanonical histogram of the order parameter is approximately flat.

The chosen order parameter interval is divided into bins of equal width with locations of the bin centers $\{ \mathcal O_i^b\}$ corresponding to the values of the order parameter at these points. The weight function is obtained for other values through a simple linear interpolation. That is,

\begin{equation}
  W(\mathcal O) = W(\mathcal O_{i}^{b}) + (\mathcal O - \mathcal O_{i}^{b})\frac{W(\mathcal O_{i+1}^{b}) - W(\mathcal O_{i}^{b})}{\mathcal O_{i+1}^{b} - \mathcal O_{i}^{b}},
\label{eq:weight-function-definition}
\end{equation}
where $i$ is such that $\mathcal O \in ( \mathcal O_{i}^{b}, \mathcal O_{i + 1}^{b} )$. What happens outside of the chosen interval depends on the end goal. We want to restrict the system to the interval and one solution is to set $W(\mathcal O)$ to some excessive value in the forbidden region. This basically guarantees that the Markov chain never leaves the interval. An alternative option is to have the weight function increase rapidly outside of the chosen interval, e.g. by replacing the quotient in Eq. \ref{eq:weight-function-definition} by some large positive (negative) constant, essentially constructing steep slopes on the right (left) side of the interval. This is convenient since if the chain is somehow initialized outside of the desired region, it will be driven there instead of just getting stuck.

\begin{figure}[htbp]
\centerline{\includegraphics[]{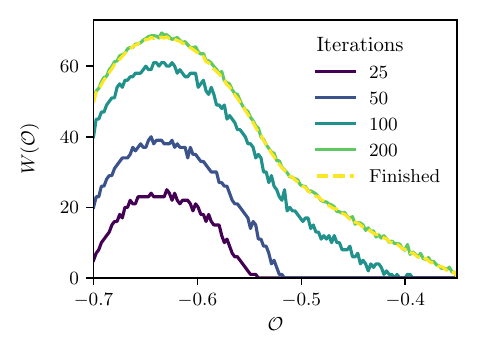}}
\caption[]{The evolution of the weight function $W$ during the iteration process. We shift $W$ such that its maximum is always at zero. This is inconsequential as only relative differences matter for the multicanonical update. \label{fig:weight-iteration}}
\end{figure}

The weight function is iterated by a pseudo Markov chain, where the weights are updated ``on the fly'' as follows:

\begin{algorithm}[H]
 $W(\mathcal O) = \mathrm{constant}$\;
 \While{$C > C_{\min}$}{
   Perform MCMC updates\;
   Measure order parameter value and find corresponding bin\;
   Add $C$ to the bin value (increasing $W$)\;
   Add the bin to the list of visited bins\;
  \If{All bins have been visited}{
   C = C / 1.5\;
   Empty the list of visited bins\;
  }
 }
\end{algorithm}
Recalling that a larger value of $W$ decreases the multicanonical probability density, it is easy to imagine why this works. The often visited regions are decreased in probability until they become unlikely enough that other regions are visited as well. $C$ is decreased each time the interval is fully covered which makes future adjustments to the weights finer. The process is considered complete when $C$ is decreased below $C_\mathrm{min}$, the value of which should be chosen such that $W$ does not change appreciably over the time it takes to visit all bins. The iteration procedure is convenient in that it requires little to no adjustments and is very easy to implement for arbitrary systems. We visualize the iteration of the weight function for the two-dimensional system in Fig. \ref{fig:weight-iteration}.

\bibliography{bibliography}
\end{document}